\documentclass[aps,twocolumn,amsmath,amssymb,showpacs,prb,superscriptaddress,unsortedaddress]{revtex4}
\usepackage{epsf}
\usepackage{graphicx}

\begin{document}

\title{Dual character of the electronic structure in YBa$_2$Cu$_4$O$_{8}$: conduction bands of CuO$_2$ planes and CuO chains}

\author{T. Kondo}
\affiliation{Ames Laboratory and Department of Physics and Astronomy, Iowa State University, Ames, IA 50011, USA}

\author{R. Khasanov}
\affiliation{Physik-Institut der Universit\"{a}t Z\"{u}rich,
Winterthurerstrasse 190, CH-8057 Z\"urich, Switzerland}

\author{J. Karpinski}
\affiliation{Laboratory for Solid State Physics ETH Z\"urich, CH-8093 Z\"urich, Switzerland}

\author{S. M. Kazakov}
\affiliation{Laboratory for Solid State Physics ETH Z\"urich, CH-8093 Z\"urich, Switzerland}

\author{N. D. Zhigadlo}
\affiliation{Laboratory for Solid State Physics ETH Z\"urich, CH-8093 Z\"urich, Switzerland}

\author{T. Ohta}
\affiliation{Advanced Light Source, Berkeley National Laboratory, Berkeley, CA 94720, USA}

\author{H.~M.~Fretwell}
\affiliation{Ames Laboratory and Department of Physics and Astronomy, Iowa State University, Ames, IA 50011, USA}

\author{A. D. Palczewski}
\affiliation{Ames Laboratory and Department of Physics and Astronomy, Iowa State University, Ames, IA 50011, USA}

\author{J. D. Koll}
\affiliation{Ames Laboratory and Department of Physics and Astronomy, Iowa State University, Ames, IA 50011, USA}

\author{J. Mesot}
\affiliation{Laboratory for Neutron Scattering, ETH Z\"urich and Paul Scherrer Institute, 5232 Villingen PSI, Switzerland}

\author{E. Rotenberg}
\affiliation{Advanced Light Source, Berkeley National Laboratory, Berkeley, CA 94720, USA}

\author{H. Keller}
\affiliation{Physik-Institut der Universit\"{a}t Z\"{u}rich,
Winterthurerstrasse 190, CH-8057 Z\"urich, Switzerland}

\author{A. Kaminski}
\affiliation{Ames Laboratory and Department of Physics and Astronomy, Iowa State University, Ames, IA 50011, USA}

\date{\today}
\begin{abstract}
We use microprobe Angle-Resolved Photoemission Spectroscopy ($\mu$ARPES) to separately investigate the electronic properties of CuO$_{2}$ planes and CuO chains in the high temperature superconductor, YBa$_2$Cu$_4$O$_{8}$. In the CuO$_2$ planes, a two dimensional (2D) electronic structure with nearly momentum independent bilayer splitting is observed. The splitting energy is 150 meV at ($\pi$,0), almost 50\% larger than in Bi$_2$Sr$_2$CaCu$_2$O$_{8+\delta}$ and the electron scattering at the Fermi level in the bonding band is about 1.5 times stronger than in the antibonding band.
The CuO chains have a quasi one dimensional (1D) electronic structure. We observe two 1D bands separated by $\sim$ 550meV: a conducting band and an insulating band with an energy gap of $\sim$240meV. We find that the conduction electrons 
are well confined within the planes and chains with a non-trivial hybridization.
\end{abstract}


\maketitle

Y-Ba-Cu-O (YBCO) is one of the most extensively studied high temperature superconductors due to its early discovery, availability of high quality single crystals and excellent technological prospects. The system is particularly interesting from a physical point of view as it combines 2D CuO$_{2}$ planes and 1D CuO chains within a single unit cell. The crystal structure leads to many interesting questions about the electronic interaction and hybridization between the planes and chains, and the role of the chains in the superconductivity.
Despite this interest, the electronic structure of YBCO has been rarely studied experimentally \cite{Campuzano,Gofron,Shen1,Shen2} with only one recent paper\cite{Borisenco_YBCO}. While these studies reveal interesting features, such as an extended van Hove singularity around ($\pi$,0)\cite{Campuzano} and bilayer band splitting in the nodal direction\cite{Borisenco_YBCO}, the handful of published studies should be contrasted with the dozens of detailed reports on Bi$_2$Sr$_2$CaCu$_2$O$_{8+\delta}$ (Bi2212). 
This is in part due to the complex termination of the cleaved surfaces in YBCO, but also to the fact that most ARPES studies of YBCO have focused on YBa$_2$Cu$_3$O$_{7}$ (Y123), which has double CuO$_{2}$ planes and a single CuO chain per unit cell. Y123 has fractional oxygen stoichiometry in the chains, which leads to complications regarding the ordering of oxygen atoms on the cleaved surface. The resulting unstable surface makes ARPES spectra difficult to interpret \cite{Shen2}. In this work, we concentrate on YBa$_{2}$Cu$_{4}$O$_{8}$ (Y124), which has double CuO chains and double CuO$_{2}$ planes per unit cell, and is naturally detwinned in contrast to Y123.  Y124 has a fixed oxygen stoichiometry and thus has a relatively simple surface state. Like its Y123 cousin\cite{Davis}, Y124 has one preferred cleavage site that lies between the BaO plane and CuO chain. After cleaving, two different types of surfaces, plane- and chain-domains are left on top of the sample. 
We report results of ARPES experiments on Y124 using a small UV beam (50-100$\mu$m). The use of $\mu$ARPES permits us to independently obtain spectra from the plane- and chain-domains and thus disentangle the electronic structure of the planes and chains.
 
Twin-free single crystals of underdoped YBa$_{2}$Cu$_{4}$O$_{8}$ with $T_c=80$K were grown by the self-flux method under high oxygen pressure\cite{Karpinski}. The ARPES experiments were carried out 
using a Scienta SES2002 and R4000 hemispherical analyzer mounted on the PGM beam line at the Synchrotron Radiation Center(SRC), Wisconsin, and beamline 7.0.1 of the Advanced Light Source, respectively. The energy resolution was 12-16 meV for photon energies ($h\nu$) between 22 eV and 33 eV, where most of the data were measured.  The angular resolution was set at $0.15^\circ$-0.30$^\circ$. All the spectra were measured at 40K. 

\begin{figure}
\includegraphics[width=3in]{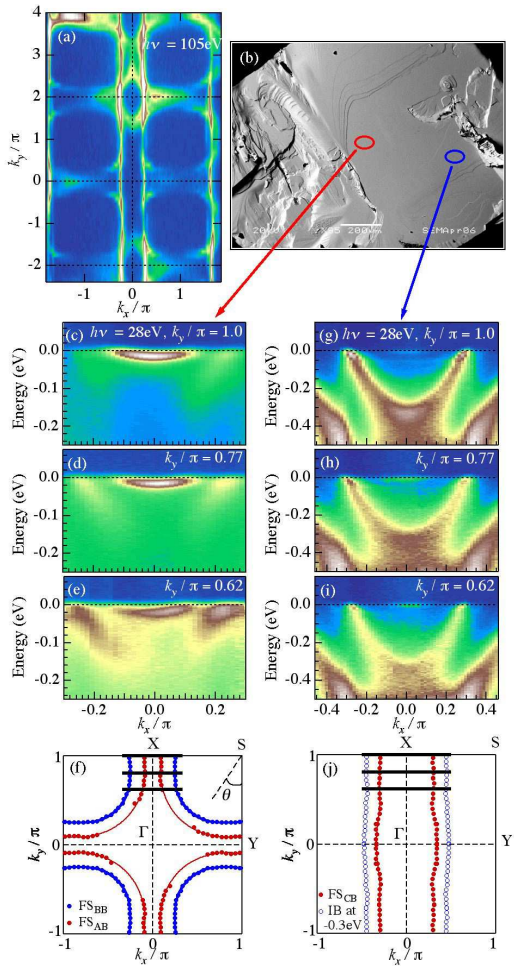}
\caption{(Color online) (a) ARPES intensity at the Fermi energy measured at $h\nu=$105eV. (b) SEM image of the sample surface after cleaving. ARPES intensity maps for (c,d,e) CuO$_2$ plane and (g,h,i) CuO chain domains obtained from area marked by red and blue oval in (b), respectively. (all measured at $h\nu=$28eV). The plane and chain data was measured along the solid lines shown in (f) and (j), respectively.
(f) Fermi surface of the antibonding band (AB) and bonding band (BB) in the planes. $\theta$ represents the Fermi angle.  (j) Fermi surface of the conducting band (CB) and the energy contour at $\varepsilon  =  - 0.3eV$ of the insulating band (IB) in the chains. Circles represent location of k$_f$ obtained from MDC  peak positions and solid lines are tight binding fit.}
\label{fig1}
\end{figure}

Figure 1(a) shows a typical ARPES intensity map at the Fermi energy for Y124 measured at $h\nu=$105eV over a wide area of the Brillouin zone. The 2D Fermi surface centered at S and the 1D Fermi surface along $\Gamma$-X are superimposed because the size of the chain and plane domains is smaller than the UV beam and  
the ARPES signal was collected from both domains. In panel (b) we show a Scanning Electron Microscope (SEM) picture of the sample surface after the ARPES experiment. The cleaved surface is not nearly as perfect as in the case of Bi2212, however it has a number of small very flat areas. By using a very small UV beam (50-100 $\mu$m) we were able to obtain ARPES data directly from these small areas and identify two types of domains with very different electronic structure. The first type - marked by a red oval in Fig. 1(b) that is scaled to represent the size of the UV beam, displayed properties consistent with the 2D electronic structure of CuO$_2$ planes, thus we will refer to it as a "plane domain". The second marked by a blue oval had the characteristic properties of the 1D CuO chains, thus we refer to it as a "chain domain".

In Fig. 1 (c)-(e), we show $\mu$ARPES data measured in the plane domain along several momentum cuts. Two features that arise from bilayer splitting are observed: a bonding band (BB) at high binding energy and an antibonding band (AB) that lies close to the Fermi energy.
From this data, we determined the Fermi surfaces of BB and AB (Fig. 1(f)). Each Fermi momentum ($\mbox{\boldmath $k$}_F$) was determined from the peak position of the Momentum Distribution Curve (MDC) at the Fermi level.
We find that both Fermi surfaces have a hole-like topology centered at S.
The area enclosed by the AB Fermi surface (FS$_{\rm{AB}}$) and the BB Fermi surface (FS$_{\rm{BB}}$) are estimated to be 72$\%$ and 51$\%$ of the Brillouin zone, respectively, indicating that the AB and BB are occupied by hole-carriers with 1.44 holes/Cu and 1.02 holes/Cu, respectively. From the average of these numbers, we determined the amount of carrier-doping ($p$) in the planes to be $p=0.23$ holes/Cu. Because it is well known that samples with a $p$ of 0.23 holes/Cu in La$_{2-x}$Sr$_x$CuO$_4$ (LSCO) are heavily overdoped, the top-most plane layer in Y124 might also be heavily overdoped and different from the bulk, which is underdoped in these samples. The spectra do not show a superconducting gap down to the lowest temperatures of 13K, which supports the idea of a heavily overdoped top-most layer.
The different doping in the bulk and surface layers is easily understood. 
Cleaving Y124 deprives the top-most CuO$_{2}$ plane of bonds to the chains, which play the role of a carrier reservoir for the planes in the bulk, and consequently the hole-concentration is larger in the top-most plane on the surface.

Figure 1 (g)-(i) shows the $\mu$ARPES chain data. Two bands are clearly seen. The higher energy band is conducting and the lower energy one is insulating. We find that the former has sharp quasiparticle peaks around the Fermi surface, and the latter has a large energy gap of $\sim$240 meV. The 1D character of both bands is revealed in Fig. 1(j), which shows the Fermi surface of the conducting band, FS$_{CB}$, and an energy contour at $\varepsilon  =  - 0.3eV$ of the insulating band. Each point was determined from the peak position of the MDC at  $\varepsilon=0$ and -0.3eV, respectively.
The signal from the chain domains consisted almost entirely of the 1D chain bands and similarly, the plane domain signal consisted almost entirely of the 2D plane bands. This indicates that the conduction electrons in Y124 are well confined within the planes and chains. Details of the electronic properties of the CuO chains are discussed in another publication \cite{RUSTEMCHAINS}

\begin{figure}
\includegraphics[width=3in]{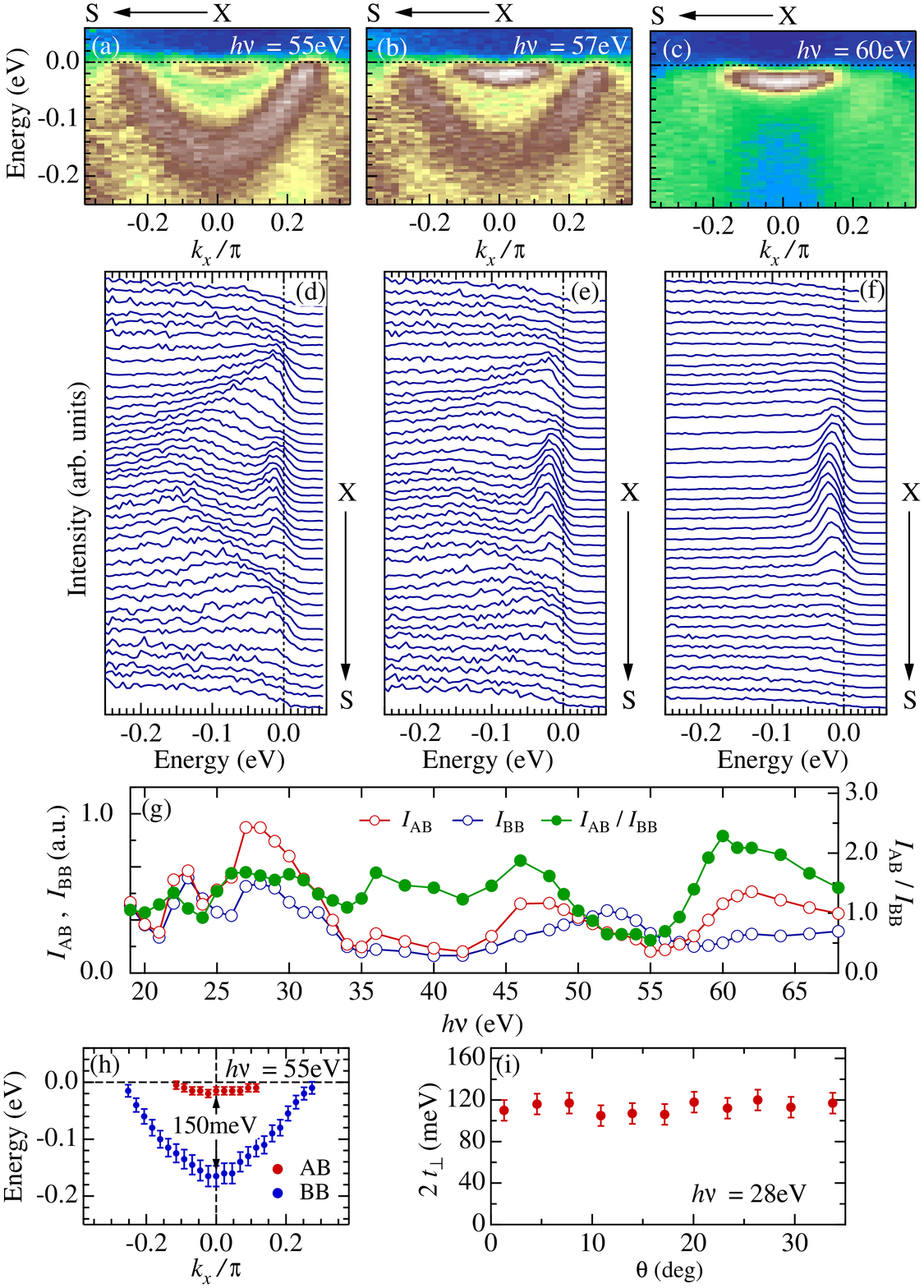}
\caption{(Color online) ARPES intensity maps  for the planes domains (a,b,c) and corresponding EDCs (d,e,f) measured along the X-S direction at $h\nu$ = 55, 57, and 60 eV. (g) Photon energy dependence of the spectral peak intensity at $\mbox{\boldmath $k$}_F$ near X in the AB ($I_{AB}$) and BB ($I_{BB}$) bands, and their ratio, $I_{AB}$/$I_{BB}$. (see text for definition of $I_{AB}$ and $I_{BB}$) (h) AB and BB dispersions determined from the EDC peak positions in (d). Arrow represents bilayer splitting energy (2$t_{^\bot}$) of 150meV at X. (i) Fermi angle $\theta$ dependence of 2$t_{^\bot}$ at $\mbox{\boldmath $k$}_F$ of AB measured at 28eV. 
}
\label{fig2}
\end{figure}

\begin{figure}
\includegraphics[width=3in]{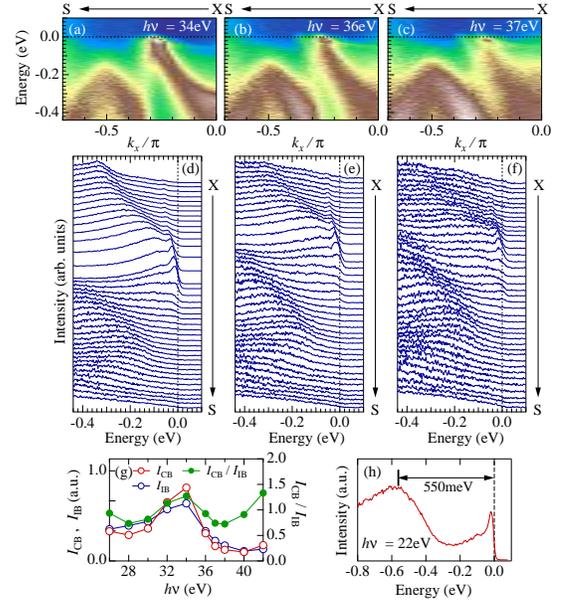}
\caption{(Color online) ARPES intensity maps of the chains domains (a,b,c) and corresponding EDCs (d,e,f) measured along X-S at $h\nu$ = 34, 36, and 37 eV. (g) Photon energy dependence of the spectral peak intensity at $\mbox{\boldmath $k$}_F$ near X for the conduction  ($I_{CB}$) and insulating ($I_{IB}$) bands and their ratio, $I_{CB}$/$I_{IB}$. (see text for definition of $I_{CB}$ and $I_{IB}$) (h) EDC at $\mbox{\boldmath $k$}_F$ near X of the conduction band measured at 22eV. Arrow represents separation energy ($\sim$550meV) between the conduction and insulating bands.  }
\label{fig3}
\end{figure}

We now investigate the photon energy dependence. 
Figure 2 (a), (b), and (c) show the ARPES intensity maps for the plane domains measured along a X-S cut at $h\nu$=55, 57 and 60eV, respectively. The corresponding energy distribution curves (EDCs) are shown in Fig. 2 (d), (e), and (f). The BB has a higher intensity at $h\nu$=55 eV, whereas the AB dominates the spectra at $h\nu$=60 eV. At $h\nu$=57 eV the intensity of both bands is equally intense. Such an anti-correlation is quite typical for bonding and antibonding bands due to the orthogonality of their wavefunctions\cite{Bansil}. 
These findings along with the result that the AB EDCs taken at 60 eV (Fig. 2(f)) show a single peak without a "peak-dip-hump" structure usually attributed to many-body effects\cite{Shen,Kaminski_kink}, lead us to conclude that the two-peak structure observed in the EDCs at 55 and 57 eV is due to bilayer band splitting.
In order to investigate the photon energy dependence of the matrix elements, we determine the spectral peak intensity of the bonding ($I_{\rm{BB}}$) and antibonding ($I_{\rm{AB}}$) bands by integrating the BB and AB EDCs at $\mbox{\boldmath $k$}_F$ over an energy range of - 70meV$\le \omega  \le$ 5meV. 
$I_{\rm{BB}}$, $I_{\rm{AB}}$ and the ratio $I_{\rm{AB}}/I_{\rm{BB}}$ measured at various photon energies are shown in Fig. 2(g). Their behavior is similar to that of the matrix elements previously reported in Bi2212 \cite{Mesot,Kordyuk,Borisenko_Bi2212}, although in the case of Y124 the overall curves are shifted to higher energy by $\sim$6eV.
From the peak positions of the EDCs in Fig. 2(d), we determine the band dispersion of the bonding and antibonding bands, and show the result in Fig. 2(h). The energy of the band splitting at S ($2t_ \bot $, where the $t_ \bot $ is the hopping integral) is estimated to be $\sim$150 meV, which is about 4 times smaller than that  obtained by band calculation, ($\sim$600 meV)\cite{Anderson,Anderson2}. The inter-layer hopping of electrons should be suppressed due to the strong electron correlation. In overdoped Bi2212\cite{Feng,Chuang}, $2t_ \bot $ at $(\pi,0)$ is reported to be $\sim$100 meV, 1.5 times smaller than in Y124.  The results indicate stronger $c$-axis coupling between double planes in Y124, and are consistent with the fact that the Y$^{3+}$ ion is smaller than the Ca$^{2+}$ ion, separating the double planes in Y124 and Bi2212, respectively. 
Figure 2(i) shows $2t_ \bot$  at various $\mbox{\boldmath $k$}_F$s of the antibonding band in Y124 as a function of Fermi angle $\theta$, defined in Fig 1(f). We find that $2t_ \bot $ is almost constant around FS$_{AB}$, which is remarkably different from overdoped Bi2212\cite{Feng,Chuang}, where $2t_ \bot $ is nearly zero at the node.
The non-zero $2t_ \bot $ at the node in Y124 is attributed to the electron orbital of the Y ion, which possesses a different symmetry from the Ca ion in Bi2212.   
Band calculations\cite{Anderson,Anderson2} suggest that the non-zero $2t_ \bot $ at the node is produced by the inter-layer hopping between the Cu $3d$ ($d_{zx}$-$d_{zx}$ $\pi$-orbitals and $d_{zy}$-$d_{zy}$ hopping) via the Y ion.
The almost constant $2t_ \bot $ around FS$_{AB}$ indicates that the inter-layer hybridization between the $\pi$-orbitals via the Y ion is relatively strong compared to that between the Cu $4s$ $\sigma$-orbitals which dominates $2t_ \bot $ around ($\pi,0$).

Figure 3 (a), (b), and (c) show typical examples of the ARPES intensity maps along X-S for the chain domains measured at $h\nu=$34, 36, and 37eV, respectively. The corresponding EDCs are shown in Fig. 3(d), (e), and (f). The conduction band with coherent peaks around the Fermi energy and the insulating band with an energy gap of $\sim$240meV at $\mbox{\boldmath $k$}_F/\pi=(-0.50,0)$ are observable. We should comment that a "peak-dip-hump" shape is seen in the spectra of the conduction band. The detailed electronic properties of the conducting chains will be discussed in a forthcoming paper. Figure 3(h) shows the EDC measured at $\mbox{\boldmath $k$}_F$ for the conduction band. The spectral peak at $\varepsilon=-0.55$eV represents the energy state of the insulating band, indicating that the conduction and insulating bands are separated by $\sim$550meV.   

We estimate the intensity of the spectral peaks in the conduction (insulating) bands, $I_{CB}$ ($I_{IB}$), by integrating the EDCs over an energy range of $ -190 \le \omega  \le$ 10meV ($ -350 \le \omega  \le$ -150meV) at $\mbox{\boldmath $k$}_F$. The resulting intensities at various photon energies are shown in Fig. 3(g). 
This data might indicate that the conduction and insulating bands arise from bilayer splitting of the double chains. However, a band calculation\cite{Muller} has estimated 2$t_{^\bot}$=36meV, which is much smaller than the separation energy ($\sim$550meV) between the conduction and insulating bands. Therefore, the conduction and insulating bands are more likely independently produced by double chains with different hole-concentrations. The different doping levels may arise because the top-most chain is deprived of the CuO$_2$ plane, which behaves as a hole accepter in the bulk. The top-most chain with no carrier doping might be half filled and produce an insulating band with a Mott gap at $k_x=\pi/2$ as observed in SrCuO$_2$\cite{Kim}, a 1D compound with non-doped double CuO chains in the unit cell.

\begin{figure}
\includegraphics[width=3.0in]{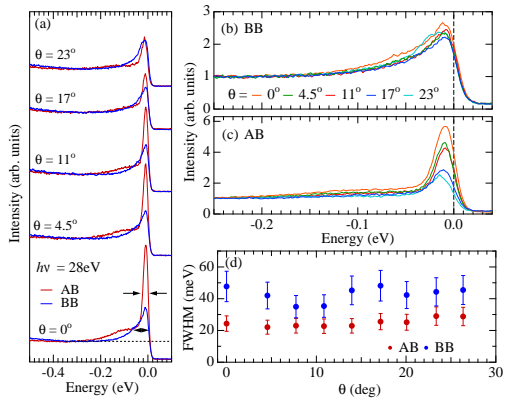}
\caption{(Color online) (a) EDCs at various $\mbox{\boldmath $k$}_F$s of the  BB and AB in the CuO$_2$ planes measured at 28eV. All spectra are normalized to the intensity at $\varepsilon  =  - 0.3eV$. Superimposed EDCs at various $\mbox{\boldmath $k$}_F$s of the (b) BB and (c) AB. (d) Fermi angle $\theta$ dependence of the full width at half maximum (FWHM) of the AB and BB EDCs (with the intensity at $\varepsilon  =  - 0.3eV$ subtracted). Estimation of the FWHM is illustrated by arrows in (a). }
\label{fig4}
\end{figure}

The lifetime of the quasiparticles is an important quantity for understanding the electron scattering mechanism, and it can be directly obtained from the peak width of the ARPES spectra\cite{Kaminski}. Figure 4(a) shows EDCs at various $\mbox{\boldmath $k$}_F$s for the BB and AB measured at $28eV$. All spectra are normalized to the intensity at $\varepsilon  =  - 0.3eV$.  
We first observe that the peak width in both BB and AB is almost independent of $\mbox{\boldmath $k$}_F$, while the matrix element effect produces a strong $\mbox{\boldmath $k$}_F$ dependence of the spectral intensity in AB. The almost constant peak width around the Fermi surface is even more evident from Fig. 4 (b) and (c), where we superimpose the EDCs at $\mbox{\boldmath $k$}_F$ for the BB and AB, respectively. We estimate the peak width (see arrows in Fig. 4(a)) from the full width at half maximum (FWHM) of the EDC with the background subtracted.The peak width in the bonding band is as about 1.5 times larger than in the antibonding, indicating that electron scattering  is stronger in the former state.
Further theoretical studies are necessary to explain this rather unexpected behavior. It may be related to fact that the wavefunctions of the bonding and antibonding states are concentrated in different parts of the unit cell, but the exact mechanism has yet to be understood \cite{ANDERSEN}.

In conclusion, we find that the conduction electrons are well confined within the CuO$_2$ planes and CuO chains with a very weak hybridization. The band of the CuO$_2$ plane is characterized by bilayer splitting, that is almost constant around the Fermi surface unlike in case of Bi2212. The lifetime of the quasiparticles in the bonding band is 1.5 times longer than in the antibonding band.


We thank O. K. Andersen and J. Schmalian for useful remarks. This work was supported by Director Office for Basic Energy Sciences, US DOE. The Ames Laboratory is operated for the US DOE by Iowa State University under Contract No. W-7405-ENG-82. The Synchrotron Radiation Center is supported by NSF DMR 9212658. ALS is operated by the U.S. DOE under Contract No. DE-AC03-76SF00098. R. K. gratefully acknowledges support of K. Alex M\"uller Foundation.

\end{document}